\documentclass[aps,prd,reprint,10pt]{revtex4-1}
\usepackage{graphicx} 
\pdfoutput=1
\begin{document}
\title{P/CP Conserving CP/P Violation \\            
 Solves Strong CP Problem}
\author{Ravi Kuchimanchi}
\date{September 28, 2010}
\email{raviaravinda@gmail.com, ravi@aidindia.org}
\begin{abstract}
We find a solution to the Strong CP problem that may be testable at the LHC and future colliders. In this solution CP is broken by parity conserving terms, while parity breaking VEVs conserve CP.  The quark mass matrix is Hermitian at the tree-level and strong CP phase is generated only in loops where CP and P violating sectors interact.  A full vector-like quark family with parity symmetric mass-terms that violate CP softly is predicted. Since no Higgs VEVs (other than the $SU(2)_L$ breaking weak-scale VEV) contribute to the masses of the new quarks, they can be naturally light, independent of  parity breaking and other high energy scales. 
\end{abstract}

\maketitle

The Strong CP Phase $\bar{\theta}$ and the Cabibbo-Kobayashi-Maskawa (CKM) phase $\delta$ are the only two phases in the standard model that can violate time reversal (T or equivalently CP) symmetry of nature. However these two phases are of very different magnitudes. Why  nature has the hierarchy $\bar{\theta} << \delta \sim 1$ is the well known Strong CP puzzle~\cite{PhysRevLett.38.1440,Nelson1984165,*PhysRevLett.53.329,PhysRevLett.56.2004,Babu:1989rb,*Barr:1991qx,*Lavoura:1996iu,Kuchimanchi:1995rp,*PhysRevLett.76.3490,*Mohapatra:1997su,*PhysRevD.65.016005}.  One key difference between the two  phases is that in addition to being CP-odd, $\bar{\theta}$ is odd under parity $(P)$ as well, and requires breaking of both P and CP symmetries to be generated.  Thus if nature respects $P$ but violates CP (or T), then it would imply that $\delta \sim 1$, $\bar{\theta} = 0$.  Indeed the electric dipole moment of the neutron, which is odd under P has not yet been observed and this is what gives the experimental~\cite{PhysRevLett.97.131801} bound $\bar{\theta} \leq 10{^{-10}}$. 

However since the weak interactions distinguish between left and right-handed light fermions, parity is not an exact symmetry and would need to be spontaneously broken. A question arises:  Is $\bar{\theta}$ spontaneously generated when parity is broken?  Since CP is not an exact symmetry, and P is broken as well, in general a sizable $\bar{\theta}$ will be generated at the tree level.    

So far there have been two work-arounds to keep $\bar{\theta}$ from being generated when $P$ is spontaneously broken  -- the first method requires adding three generations of mirror  families so that the $\bar{\theta}$ generated cancels between the normal and mirror families~\cite{Babu:1989rb,*Barr:1991qx,*Lavoura:1996iu}.    The other invokes supersymmetry~\cite{Kuchimanchi:1995rp,*PhysRevLett.76.3490,*Mohapatra:1997su,*PhysRevD.65.016005} but these models are challenging due to their complicated vacuum structure~\cite{Kuchimanchi:1993jg,*Kuchimanchi:1995vk,*huitu-1995-344,*aulakh-1997-79}. Likewise in models where CP symmetry has been used to set $\bar{\theta}$ to zero~\cite{Nelson1984165,*PhysRevLett.53.329,Bento:1990wv,*Bento:1991ez,Branco:2003rt,Glashow:2001yz}, it is spontaneously  generated at  the tree level when CP is broken unless other symmetries are imposed.     The basic question remains, how are P and CP violated in nature without generating sizable $\bar{\theta}$?  

In this work we probe this question. First we note that parity violation would necessarily involve the Higgs fields since the left-right interactions are gauged and need to be broken. However there is no such fundamental need to invoke spinless Higgs fields for CP violation. Since $\bar{\theta}$ requires both P and CP violation, if the Higgs VEVs that break parity respect CP, they would potentially not contribute to $\bar{\theta}$ directly. The Higgs VEVs would naturally respect CP,  if the Higgs does not directly participate in CP violation.  This then implies that CP violation would have to come softly from direct dimension-3 quark mass terms, which are only possible if there are  vectorlike quarks. Since these terms do not couple to the Higgs, they respect P   even while violating CP, and so they would not generate $\bar{\theta}$ at the tree-level either.

Thus we envision that P violation occurs due to CP symmetric VEVs, while CP violation is caused in nature by P symmetric mass terms involving new vectorlike quarks.  Either CP or P continues to protect $\bar{\theta}$ though both P and CP are broken.     As seen in the remainder of the paper Strong CP phase is not generated at the tree level and appears at the one-loop level through small off-diagonal couplings between the existing families and the fourth family. There is no triplication of existing families by mirror families.

With the notable exception of the possibility of a massless up quark~\cite{PhysRevLett.56.2004}, spin zero fields have hitherto played a special role in attempts to solve the strong CP puzzle.   In the Peccei-Quinn mechanism~\cite{PhysRevLett.38.1440} the spin-zero axion dynamically rotates $\bar{\theta}$ to zero. In Nelson-Barr mechanism~\cite{Nelson1984165,*PhysRevLett.53.329,Bento:1990wv,*Bento:1991ez,Branco:2003rt} they  are necessary to generate CP phases without inducing $\bar{\theta}$ at the tree level and their masses have quadratic divergence that links them to the GUT scale. Likewise in the more recent split fermion approach~\cite{Harnik:2004su} in 5D where  P in the bulk and CP in the brane protect $\bar{\theta}$ at the tree level, it is generated in loops with the scalar mediating CP violation.   However in our model the CP violating phases can occur in parity symmetric mass terms involving vector-like quarks without the need for participation of spin-zero Higgs fields.   As a result no Higgs VEVs break the automatic symmetry protecting the masses of the new quarks.  Hence these can be naturally light including being accessible to the LHC.

\section{Left-Right Model With Vector Like Iso-Doublet Quark Family}
We consider the Left-Right symmetric model~\cite{PhysRevD.10.275,*PhysRevD.11.566} given by $SU(3)_C \times SU(2)_L \times SU(2)_R \times U(1)_{B-L} \times P$ (for some recent articles please see~\cite{Zhang:2007da,Maiezza:2010ic} and references therein).   

In addition to the usual 3 light chiral color triplet quark families $Q_{iL}$ and $Q_{iR}$, which are doublets of $SU(2)_L$ and $SU(2)_R$ respectively,  we add a fourth  vector like quark family, whose left and right handed components $Q_{4L}$ and $Q'_{R}$ are both $SU(2)_L$ doublets.    Due to parity there is also the corresponding $SU(2)_R$ doublet, $SU(2)_L$ singlet (iso-singlet) vector-like family with components:  $Q_{4R}$, $Q'_{L}$.   We can think of these as the fourth iso-doublet and fifth  iso-singlet vector-like families or preferably as the fourth \emph{full}  vector-like quark family with the normal $4^{th}$ chiral and primed flipped-chiral components. Note that such families have been considered occasionally in the past for different reasons~\cite{Babu:1994pd,Pirogov:1999gc}.

Anomaly cancellations require the usual 3  color-singlet chiral light lepton families, while vector-like lepton families are not required and  may or may not exist.  In this work we study mainly the quark sector and do not mention the leptons explicitly. 

Parity exchanges the $SU(2)_L$ and $SU(2)_R$ gauge bosons,  and  $Q_{iL} \leftrightarrow Q_{iR}, {Q'}_{L} \leftrightarrow {Q'}_{R}$, where here after we assume $i$ runs from $1$ to $4$ to include the usual 3 light chiral and the $4^{th}$ normal chiral component of the full vector like family.  


Note that $\bar{\theta} = \theta_{QCD} + \arg \det \left( M_u M_d \right)$ where $\theta_{QCD}$ is the coefficient of the $F \tilde{F}$ term in the QCD Lagrangian, and $M_u$ and $M_d$ are the up and down quark mass matrices.  Being a winding number $F \tilde{F}$ is odd under parity and  since the Lagrangian in our model is parity invariant, $\theta_{QCD}$ vanishes. 

We will first show that if CP is respected by the quartic terms of the Lagrangian, then VEVs that break parity (which is a good symmetry of the Lagrangian) conserve CP. We will allow for CP to be broken softly by parity symmetric dimension 3 quark mass terms involving vector-like quarks. This leads to mass matrices $M_u$ and $M_d$ being Hermitian and the strong CP phase is zero at tree level. In section~\ref{sec:oneloop} we estimate radiative corrections and in section~\ref{sec:SPCP} we generalize the model so as to have spontaneous breaking of CP and/or P. No other symmetries are imposed and our treatment applies to the most general Lagrangian. 

\noindent \subsection{CP Conserving P Violation} In the minimal left-right symmetric model, parity is spontaneously broken when the neutral component of $SU(2)_R$ Higgs triplet $\Delta_R$ (or doublet $\chi_R$), picks up a VEV $v_R  > v_L$, where $v_L$ is the $SU(2)_L$ breaking weak scale.   Note that if the coefficient $\lambda$ is positive, quartic terms in the Higgs potential such as $\lambda ( Tr \Delta_R^\dagger \Delta_R Tr \Delta_L^\dagger \Delta_L)$ are minimized when $\left<\Delta_L\right>=0$. Because of such terms $\Delta_R$ alone picks a large VEV. Thus parity is spontaneously broken.  Note that without loss of generality we can use $SU(2)_R$ invariance to align the VEV of $\Delta_R$  to be real or CP conserving. 

As is usual in the minimal left-right symmetric model, the $B-L=0$, $SU(2)_L \times SU(2)_R$ bi-doublet Higgs $\phi$ picks up $SU(2)_L$ breaking VEVs  and provides masses to the quarks and leptons. By convention we use $SU(2)_L$ invariance to set one of the two VEVs of $\phi$ to be real.  We now show that if CP is conserved by the quartic terms, then all terms involving the bi-doublet in the Lagrangian are real and therefore both its VEVs are real and CP conserving. 

The Yukawa terms involving the bidoublets are: 
\begin{widetext}
 {
\begin{equation}
\underline{h}_{ij} \ \bar{Q}_{iL} \ \phi \ Q_{jR} \ + \  \tilde{\underline{h}}_{ij} \ \bar{Q}_{iL} \ \tilde{\phi} \ Q_{jR} \  + 
\underline{h}' \  {\bar{Q'}}_{L} \ \phi^\dagger \ {Q'}_{R} \ + \ \tilde{\underline{h}}' \ {\bar{Q'}}_{L} \ {\tilde{\phi}}^{\dagger} \ {Q'}_{R} \ + \ h.c,
\label{eqn:yukawa}  
\end{equation}}
\end{widetext}
\noindent Under Parity $\phi \leftrightarrow \phi^{\dagger}$ and we have used the notation $\tilde{\phi}$  for $\tau_2 \phi^{\star} \tau_2$.  Invariance of Yukawa terms under parity implies that  {$\underline{h}_{ij} = {\underline{h}^\star_{ji}}$,  $\tilde{\underline{h}}_{ij} = {\tilde{\underline{h}}^\star_{ji}}, \underline{h}' = \underline{h}'^{\star}$}  and \itshape $\tilde{\underline{h}}'=\tilde{\underline{h}}'^{\star}$ \normalfont. Moreover CP invariance of the quartic terms implies that they are all real. Thus all Yukawa couplings are real and Yukawa matrices are real symmetric.   Likewise, quartic terms of the Higgs potential involving $\phi$, and $\Delta_{L, R}$ or $\chi_{L, R}$ are real as well,  since CP is broken softly by dimension 3 quark mass terms and is conserved by the quartic terms.  

Note that for clarity we are using the underline in the notation  {$\underline{h}$} to indicate real Yukawa couplings in the symmetry basis.

It is easy to check that despite soft CP violation, the quadratic and cubic terms of the Higgs fields cannot have complex coefficients as they are real due to Parity.  Thus the coefficients of terms such as ${\tilde{\phi}}^\dagger \phi$ and $\chi_L^\dagger  \phi \chi_R$ and $\chi_L^\dagger \tilde{\phi} \chi_R$ are real.  This is the case even if we have more than one bi-doublet.

Thus in our model there is no CP violating term that involves any of the Higgses.  Thus all Higgses, $\Delta_{L,R}$ and $\phi$ pick up real CP conserving VEVs.  $\phi$ picks up the VEV $diag \left\{\kappa, \kappa'\right\}$ where $\kappa$ and $\kappa'$ are real and ${v_L}^2 \sim \kappa^2 + \kappa'^2$.  Note that in our notation $v_L$ and $v_R$ denote the left and right symmetry breaking scales and the neutral component of $\Delta_L$ picks up a small real VEV that we denote by $\delta_L \sim v_L^2/v_R$

\noindent \subsection{P Symmetric CP Violation}
\label{sec:CPV}
\noindent CP is broken softly by complex phases in parity symmetric mass terms:  
\begin{equation}
M_i \ \bar{Q}_{iL}  {Q'}_{R} \ + \ M_i^{\star} \  \bar{{Q'}}_{L}  Q_{iR} \ + \ h.c
\label{eq:massterms}
\end{equation}
where the sum over repeated index i goes from 1 to 4. 

Note that the CP violating coefficients $M_i$ and $M_i^{\star}$ occur together ensuring that the sum of the above two terms is parity invariant. These terms give masses to the new vector-like quarks and can be naturally small.  This is because setting $M_i$ to zero  restores the automatic global Baryon number symmetry of the primed (flipped-chiral) quarks, ${Q'}_{L,R} \rightarrow e^{i\alpha}{Q'}_{L,R}$, that the rest of the terms in the Lagrangian possess. The VEVs of Higgs fields discussed earlier do not break this symmetry.   Thus $M_i$ can be naturally small in the sense of 't Hooft~\cite{tHooft}, including being at scales accessible to the LHC.

\subsection{No Strong CP Phase At Tree Level}

After electro-weak symmetry breaking the up quark masses are given by:
\begin{math}
\left({{\bar{u}}_L, \bar{u}'_{L}}\right) \ M_u \ 
\left(\begin{array}{c}
{u}_R  \\ 
u'_{R}  \ 
\end{array}\right)
\end{math}
with
 {
\begin{equation}
M_u =
\left(\begin{array}{cccc|l}      
\ \  & \ \  & \ \   & \  \ &  M_1 \\       
\ \  & \underline{H}_u v_L  & \ \ & \ \ & M_2 \\
\ \ & \ \   & \ \  & \ \  & M_3 \\
\ \  & \ \  & \ \  & \ \  & M_4 \\
\hline
M_1^\star & M_2^\star & M_3^\star & M_4^\star &  {\underline{h}'}_u v_L 
\end{array}\right)
\label{eq:massmatrix}
\end{equation}}
 where $\bar{u}_L$ is the $1 \times 4$ row vector $\left(\bar{u}_{1L},..., \bar{u}_{4L}\right)$ and {$\underline{H}_u v_L$} is a $4 \times 4$ real symmetric matrix with components  $\underline{h}^u_{ij} v_L = \underline{h}_{ij} \kappa + \tilde{\underline{h}}_{ij} \kappa'$.  Likewise  $\underline{h}'_u v_L = \underline{h}' \kappa + \tilde{\underline{h}}' \kappa'$ is real.  The down quark mass matrix $M_d$ is similar to the up quark mass matrix with  {$\underline{H}_u \rightarrow \underline{H}_d, \underline{h}'_u \rightarrow \underline{h}'_d$} and $\kappa \leftrightarrow \kappa'$ in equations determining them.  Note that in general  {$\underline{H}_u$} and  {$\underline{H}_d$} do not commute.  

Since the up and down quark mass matrices are Hermitian, $\arg \det \left(M_u M_d\right) = 0$ and there is no strong CP phase at the tree level. Note that the matrices are not in the form of Nelson-Barr~\cite{Nelson1984165,*PhysRevLett.53.329,Bento:1990wv,*Bento:1991ez,Branco:2003rt} or Glashow's~\cite{Glashow:2001yz} model where complex phases are multiplied by zeros in the mass matrix to ensure that the determinant is real.  Instead all matrix elements can be present and the determinant is real due to hermiticity. 

To see that there is a non-trivial CKM phase we note that had  {$\underline{H}_u$} and  {$\underline{H}_d$} commuted, then they could be simultaneously diagonalized by a unitary transformation in the $4 \times 4$ sub-space. On diagonalizing the $4 \times 4$ sector, the resultant phases in the 5th column (and row) of the mass matrix can all be rotated away for the up and down sector quarks, thereby implying no weak CP violation. However since in general  {$\left[ \underline{H}_u,  \underline{H}_d \right] \neq 0$} the phases in $M_i$ cannot all be rotated away and therefore we have weak CKM CP violation without Strong CP violation.

We show through an example in the appendix~\ref{sec:appendix} that on diagonalization of the mass matrix a non-trivial CKM phase is generated that depends on the mass ratios $M_i/M$ and yukawa couplings, where M is the mass eigenvalue of the vector-like quarks.  Since it depends on  ratios of direct mass terms which are all protected by the same automatic symmetry it is not diluted even for  $M >> v_L$. 

In the Nelson-Barr models the CP phase is generated by ratios that involve Higgs VEVs that break CP and direct quark mass terms, and these two distinct mass scales  are  \emph{assumed} to be similar~\cite{Branco:2003rt} so as to generate the CKM CP phase. We do not need to make such an assumption as our model works without the involvement of Higgs VEVs. 

Note that there are totally 3 CP violating phases since $M_4$ can be made real through a choice of basis. One combination, the CKM phase is unsuppressed, while effect of others for weak CP violation is attenuated by powers of $v_L/M$.  

\noindent\subsection{Nearly Degenerate Masses of Vector-Like Fourth and Fifth Family Quarks}
In the absence of $v_L$ the  matrix $M_u$ (and likewise $M_d$) has 3 zero eigenvalues corresponding to the 3 light chiral quarks.  The 2 non-zero eigenvalues ($\pm M$) are equal in magnitude and correspond to the masses of the heavy iso-doublet and isosinglet up  quarks: $M =  \sqrt{\sum \left|M_i\right|^2}$.   Note that in principle it is possible to break parity spontaneously in such a way that the masses are different.   For example  a parity odd singlet or  $B-L=0$, $SU(2)_R$ triplet (with its $SU(2)_L$ partner) can be  introduced  and given a VEV to split these masses. But such a VEV would also make the complex mass terms non-Hermitian thereby generating a Strong CP phase.  Thus $\bar{\theta} \sim 0$ tells us that the fourth and fifth up quark masses (and likewise for the corresponding down quarks)  are equal at the scale M and parity is only broken at scale $v_R$ through VEVs that do not split them at the tree level. 

Since ${v_R} > v_L$ the masses of iso-doublet and isosinglet quarks will receive slightly different one-loop radiative corrections from $W_L$ and  $W_R$ respectively and accordingly they will be split by $\sim ({g^2}/{16{\pi^2}}) M$ times some logarithmic factor.   Like wise masses are split by $\sim h  v_L$ with the breaking of the electro-weak symmetry where $h$ are the typical Yukawa couplings. The four heavy quarks -- two up and two down -- all having the same large mass M split slightly is a prediction of the model.   As we saw before, $M$ is  protected from quadratic divergences and it may be naturally accessible at current or future colliders. 

With experiments showing that neutrinos have very small mass-squared differences ($ < 10^{-3} eV^2$), it is possible that the parity breaking scale $v_R$ is very high so that $m_{\nu} \sim {h_\nu}^2 v_L^2/v_R \sim 0.1 eV$.   However the right-handed scale does not enter into the quark mass matrix in our model. Thus $M$ is independent of $v_R$ and can be low even if $v_R$ is very high including being at the GUT scale. The near degeneracy of the vector-like quark masses at scale $M$ as predicted in this model can be an exciting indicator of left-right symmetry and the strong CP solution, even if the parity breaking scale is much higher. 

\noindent \section{Radiative Corrections to $\bar{\theta}$} 
\label{sec:oneloop}
Since P and CP violating terms will interact with one another in loops we expect $\bar{\theta}$ to be generated radiatively. Without loss of generality we can choose a flavour basis via a common $4 \times 4$ unitary transformation $U$  on the up and down sectors that depends on the phases in $M_i/M$, such that $U^{\dagger}_{ij}M_j = \delta_{i4} M$  where $\delta_{i4} = 0$ for $i < 4$ and $\delta_{44}=1$.  In the transformed  basis $H_u$ and $H_d$ are hermitian matrices with complex phases as they are obtained from real symmetric matrices by  $H_{u,d} = U^\dagger \underline{H}_{u,d} U$. Correspondingly the mass matrices in~(\ref{eq:massmatrix}) transform as  $M_{u,d} \rightarrow \left(\begin{array}{cc} U^\dagger & 0 \\ 0 & 1 \end{array}\right) M_{u,d} \left(\begin{array}{cc} U & 0 \\ 0 & 1 \end{array}\right)$. We can now apply a common unitary rotation in the light $3 \times 3$ sector to diagonalize the $3 \times 3$ subspace of $H_d$ and rotate away all phases in the $4^{th}$ row (and column) so that it is real symmetric. In this basis $H_u$ is a $4 \times 4$ hermitian matrix with complex coefficients.  Such a basis has been used in the past to analyze vector-like quarks~\cite{Pirogov:1999gc}.  Writing explicitly the matrix elements $h_{dd}$ and $h_{uc}$ etc of $H_d$ and $H_u$ respectively,  the up and down mass matrices look as follows in this physical basis:
\noindent
\begin{equation}
M_d =
\left(\begin{array}{cccc|l}      
 h_{dd} v_L &  0  & 0   & h_{d4} v_L&  0 \\       
 0 & h_{ss} v_L & 0 & h_{s4}v_L  & 0 \\
 0& 0  & h_{bb} v_L &  h_{b4} v_L & 0 \\
h_{d4} v_L & h_{s4}v_L  &  h_{b4} v_L & h^{d}_{44} v_L &  M\\
\hline
0 & 0 & 0 & M &  h'_d v_L 
\end{array}\right)
\label{eq:Md_diag_basis}
\end{equation}
\smallskip
\newline
\noindent
\begin{equation}
M_u =
\left(\begin{array}{cccc|l}      
h_{uu}v_L  & h_{uc}v_L  & h_{ut}v_L   & h_{u4} v_L &  0 \\       
h^\star_{uc}v_L  & h_{cc}v_L  & h_{ct}v_L & h_{c4}v_L & 0 \\
h^\star_{ut}v_L & h^\star_{ct}v_L   & h_{tt} v_L  & h_{t4}v_L  & 0 \\
h^\star_{u4}v_L & h^\star_{c4}v_L  & h^\star_{t4}v_L  & h^u_{44}v_L  &  M\\
\hline
0 & 0 & 0 & M &  h'_u v_L 
\end{array}\right)
\label{eq:Mu_Md_diag_basis}
\end{equation}
\newline
where $M_d$ is real symmetric and $M_u$ is Hermitian with complex coefficients $h^{u}_{ji} = h^{u\star}_{ij}$ Note that we drop the superscript $u$ (or $d$) whenever it is obvious by the subscripts that a Yukawa coupling belongs to the up (or down) quark matrix.  

We denote by $\delta M_{d_{ij}}$ the radiative corrections to the $i^{th}$ row and $j^{th}$ column of $M_d$ and evaluate the determinant to the lowest order in $\delta M_d$ and find:
\begin{eqnarray}
\arg \det  (M_d + \delta M_d) = & \nonumber \\  Im \left(  \sum_{i=d,s,b} {\frac{\delta M_{d_{ii}}}{h^d_{ii}v_L}} +  \sum_{i=d,s,b,4} \frac{h^d_{i4}}{h^d_{ii}} \frac{\left( \delta{M_{d}}_{i5} + \delta{M_{d}}_{5i}\right)}{M}\right) & 
\label{eq:det}
\end{eqnarray}
To arrive at the above  form we have set $h'_d = h'_u = 0$ in~(\ref{eq:Md_diag_basis}) and~(\ref{eq:Mu_Md_diag_basis})  as this can be obtained by imposing the softly broken chiral symmetry $Q'_L \rightarrow e^{i\beta} Q'_L$ on the Yukawa terms of equation~(\ref{eqn:yukawa}).  The rest of the terms in $\delta M_{d_{ij}}$, for example with $i \neq j \leq 4$, do not contribute to the determinant if $h'_d=0$.  Note that this simplification is just to ease the calculations and in general no symmetries need to be imposed.

Note that $\delta {M_u}$ also contributes to $\bar{\theta}$ through an equation similar to~(\ref{eq:det}) with $d \rightarrow u$ evaluated in the basis where the light $3 \times 3$ sector of $M_u$ is diagonal.  We now calculate the radiative corrections $\delta{M_d}$ to equation~(\ref{eq:Md_diag_basis}) and towards the end of this section discuss the result for $\delta{M_u}$ as well.

CP violating phases in $M_u$ can provide corrections to $M_d$ through the charged weak current which violates parity.     
\begin{figure}[ht]
\includegraphics[width=55mm]{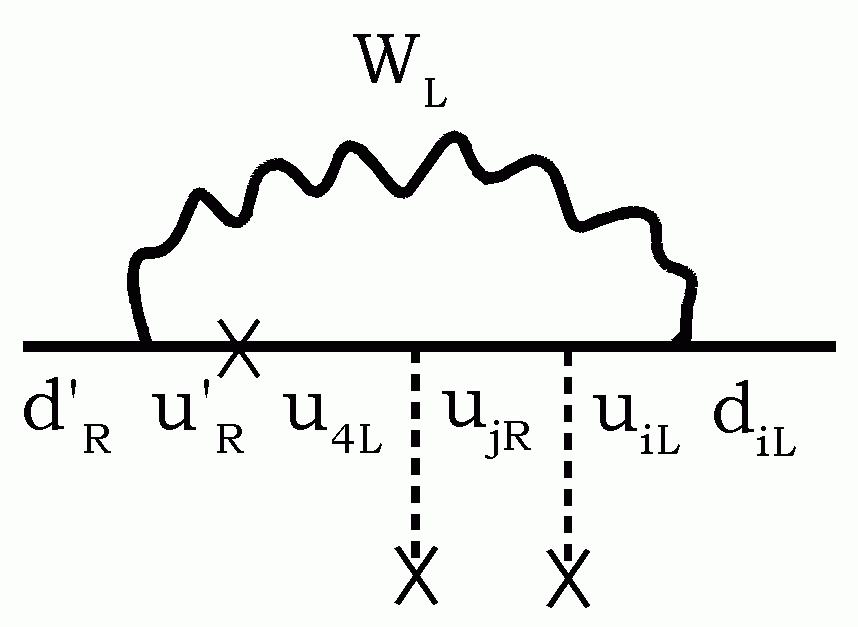}
\caption{One loop contribution of charged current to $\bar{\theta}$}
\label{oneloopW}
\end{figure}  
The one-loop contribution in Figure~\ref{oneloopW} is suppressed by the mass of the heavy quarks and is
\begin{equation}
\delta {M_{d}}_{i5} \sim \frac{g^2 h^u_{ij}h^{u}_{j4}}{16 \pi^2} \frac{{v_L}^2}{M}
\label{eq:deltaMd}
\end{equation}  
where the sum over the repeated index $j$ is implied. By substituting the result for $\delta {M_d}_{i5}$ into  equation~(\ref{eq:det}) we  estimate
the contribution of figure~\ref{oneloopW} to be:

\begin{equation}
\bar{\theta}_{W_L \ loop} \sim  \left({\frac{g^2}{16 \pi^2}}\right)   \ Im \left(\frac{{h^{d}_{i4}}{h^{u}_{4j}}h^{u}_{ji}}{h^d_{ii}}\right) \left[{\frac{v_L}{M}}\right]^2
\label{eq:top_yukawa}
\end{equation}
 
We now turn to the Higgs.  It is known that for the case where the VEVs for $\phi$ and $\Delta_{L,R}$ (or $\chi_{L,R}$) are all real that in the decoupling limit of $v_R >> v_L$ the natural theory at low energies is  the  standard model with a single Higgs doublet. On the other hand if there is spontaneous CP violation then we would have two Higgs doublets in this limit~\cite{PhysRevD.65.095003,*PhysRevD.44.837}. Since all the VEVs in our model are real, there is only the standard model Higgs at the low energy.  The neutral Higgs of the standard model alone cannot generate the strong CP phase in loops since in our basis where $M_d$ is real, the Yukawa matrix for the down sector will  also be real for this Higgs.   

In general we should also consider the effect of the second Higgs doublet that comes from the bi-doublet. Note that even after $\Delta_R$ and $\phi$ pick up  VEVs, the Yukawa terms in equation~\ref{eqn:yukawa} with $\phi$ replaced by $\left<\phi\right> + \phi$ remain invariant under $\phi \rightarrow \phi^\dagger, Q_{iL,R} \rightarrow Q_{iR,L}, Q'_{L,R} \rightarrow Q'_{R,L}$. This is because $\left< \phi \right>$ is P conserving.  However the Higgs potential needs to be diagonalized to obtain the physical Higgs mass eigenstates.  Since parity is broken spontaneously in general the Higgs mass eigenstates may not all be parity eigenstates. When $\phi$ is expressed in terms of mass eigenstates, the resulting Yukawa terms for mass eigenstates that do not have a definite parity can generate a strong CP Phase at one-loop.  While the Yukawa interactions for parity eigenstates will not generate $\bar{\theta}$ in one loop. The eigenstates of the Higgs potential of the left-right symmetric model are well-studied and we note from~\cite{Duka:1999uc} that all the charge-zero Higgs mass eigenstates that participate in Yukawa interaction (involving both standard model doublets) are also parity eigenstates. Thus they do not contribute to the strong CP phase in one loop.
\begin{figure}[ht]  
\includegraphics[width=55mm]{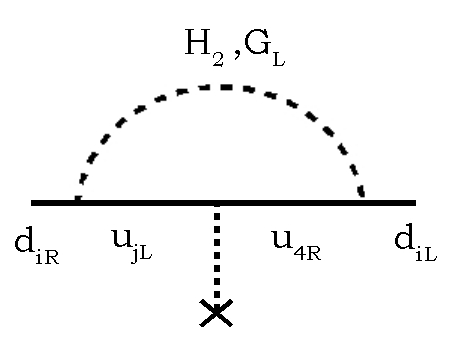}
\caption{Contribution due to Higgs mass eigenstates that do not have a definite parity}
\label{oneloophiggscharged}
\end{figure}

The charged component of the second doublet however is a Higgs mass eigenstate that does not have a definite parity. Its Yukawa interaction terms for 3 families are well known~\cite{Duka:1999uc, Zhang:2007da} and translate to our  case with 4 families.  The Yukawa terms of the charged second Higgs doublet that contribute  in one-loop to $\bar{\theta}$ are:
\begin{equation}
\mathcal{L}_Y = \bar{u}_{iL}h^{u}_{ij} d_{jR} H^+_2 -\bar{u}_{iR}h^{d}_{ij} d_{jL} H^+_2 + h.c. +...
\label{eq:chargedhiggs}
\end{equation}
where we have not indicated terms whose contribution vanishes. The would-be Goldstone mode $G^+_L$  also has Yukawa couplings and we note that they can be obtained from the first two terms in~(\ref{eq:chargedhiggs}) with $L \leftrightarrow R, H^+_2 \leftrightarrow G^+_L$. Due to the left-right symmetric nature of the substitution, the divergent contribution to $Im \ \delta {M_d}_{ii}$ from Figure~\ref{oneloophiggscharged} cancels between $H^+_2$ and $G^+_L$ in the Feynman gauge and a finite contribution to $\bar{\theta}$ emerges that depends on the diagonal elements of the product $H_d H_u H_u$. 

Since the mass of $H^+_2$ is $\sim v_R$ and that of the 4th quark in Figure~\ref{oneloophiggscharged} is $\sim M$, we recognize two physically distinct cases for estimating $\bar{\theta}$: If $M > v_R$ then the CP breaking vector-like quark mass terms are heavier than the scale of P breaking.  On the other hand if $M < v_R$ they are lighter than the P breaking scale.

\subsection{Case 1: $M > v_R$}

In this case the only effect of the vector-like quarks is to induce the CKM phase when the mass matrix is diagonalized. When we integrate the heavy vector-like quarks out  since P is broken at a lower scale, it protects $\bar{\theta}$ from being generated.  Thus we expect in this case that $\bar{\theta}$ from figure~\ref{oneloophiggscharged} is proportional to powers of $v_R /M$ and vanishes as $M \rightarrow \infty$. 

Writing explicitly the diagonal elements of the product $H_d H_u H_u$ we have
\begin{equation}
\bar{\theta}_{Higgs \ M > v_R} \sim  \left({\frac{1}{16 \pi^2}}\right)   \ Im \left(\frac{{h^{d}_{i4}}{h^{u}_{4j}}h^{u}_{ji}}{h^d_{ii}}\right) \left[{\frac{v_R}{M}}\right]^2
\label{eq:theta_higgs}
\end{equation} 
To obtain the above note that we are working in the basis where $H_d$ has the form in the $4 \times 4$ sector of $M_d$ in~(\ref{eq:Md_diag_basis}), and we have dropped the terms that are all real. The denominator $h^d_{ii}$ comes from the first term in equation~(\ref{eq:det}).
 
Also note that the contribution to $\bar{\theta}$ from $\delta M_{i5}$ through equation~(\ref{eq:top_yukawa}) for the case $v_R < M$ is canceled by a parity symmetric contribution to $\delta M_{5i}$ with $W_R$ gauge boson running in the loop of Figure~\ref{oneloopW} in place of $W_L$ and $L \leftrightarrow R$. 
 
This solution for the strong CP problem in left-right models works with either $\Delta_R$ or $\chi_R$ as Higgs choices to break parity.  Note that previous solutions have relied on $\chi_R$ as well as triplication of quarks by mirror quarks and do not work with $\Delta_R$ which is phenomenologically  motivated by the see-saw mechanism for neutrino masses. 

\subsection{Case 2: $ M < v_R$}

The case $M < v_R$ is phenomenologically the most interesting from the point of view of testing this model at LHC.  As we have seen, no Higgs VEVs (other than the $SU(2)_L$ breaking weak-scale VEVs) contribute to the mass of the vector-like quarks and they can be naturally light. Since parity breaks at a higher scale there is no suppression of the one-loop contribution from the Higgs and we have:
\begin{equation}
\bar{\theta}_{Higgs \ M < v_R} \sim \sum_{\stackrel{i =1 \ to \ 3}{j = 1 \ to \ 4}} \left({\frac{1}{16 \pi^2}}\right)   \ Im \left(\frac{{h^{d}_{i4}}{h^{u}_{4j}}h^{u}_{ji}}{h^d_{ii}}\right)
\label{eq:theta_higgs_lowM}
\end{equation}  
The above relation together with $\bar{\theta} \leq 10^{-10}$  implies that some of the Yukawa couplings involving the fourth family must be $\sim 10^{- (4 \ to \ 5)}$.   Given that some Yukawa couplings of the up and down quarks and that of the electron are $\sim 10^{-5}$ this is not a severe  constraint. 

Since we know from the first three families that there is a hierarchy in Yukawa couplings, it is possible that the fourth family is heavy only due to its vector-like nature rather than due to a large Yukawa. Its Yukawa couplings can fall anywhere in the hierarchy including being like those of the first family or below it even. 

Also note that the contribution to $\bar{\theta}$ from $W_L$ loop through equation~(\ref{eq:top_yukawa}) is not cancelled by $W_R$ which is much heavier for the case $v_R > M$.  However since equation~(\ref{eq:theta_higgs_lowM}) is a stronger constraint than equation~(\ref{eq:top_yukawa}) it implies that the mass $M$ of the vector-like quark is not bounded from below  and it can be as low as $3 v_L$ or in the TeV scale.  

Equation~(\ref{eq:theta_higgs_lowM}) refers to the couplings in the physical basis given by~(\ref{eq:Md_diag_basis}),(\ref{eq:Mu_Md_diag_basis}).  These can be expressed in terms of the symmetry basis Yukawa couplings in~(\ref{eq:massmatrix}) and mass ratios $M_i/M$ of the direct mass terms of the vector-like quarks through a unitary transformation. Thus  equation~(\ref{eq:theta_higgs_lowM}) will imply that the some of the symmetry basis couplings as well as mass ratios will be correspondingly small. In appendix~\ref{sec:appendix} an example has been discussed where we see how the model can generate the CKM phase while some Yukawa couplings are small to ensure that $\bar{\theta}$ is within bounds.  

It is worth looking at possible flavour symmetries that would explain the hierarchy of Yukawa couplings observed in nature.  If such symmetries are imposed without any new Higgs VEVs breaking the automatic symmetry protecting the masses of the vector-like quarks, then these will continue to remain naturally light. 

We have calculated the contribution to $\bar{\theta}$ from $\delta M_d$ using the basis where the light $3 \times 3$ sector of $M_d$ is diagonal.  The results can also be expressed in an invariant way so that the calculation can be done in any basis corresponding to unitary transformations $U'$ in the light $3 \times 3$ sector (with $U'_{i4} = U'_{4i} = \delta_{i4}$). This will in turn help us calculate the contribution due to $\delta M_u$.  Note for example that equation~(\ref{eq:theta_higgs_lowM}) can be expressed equivalently as 
\begin{equation}
\bar{\theta}_{Higgs \ M < v_R} \sim \sum_{\stackrel{i,l =1 \ to \ 3}{j,k = 1 \ to \ 4}} \left(\frac{1}{16 \pi^2}\right)Im \left({h^{d}_{ik}}{h^{u}_{kj}}h^{u}_{jl}{\left(h_{3 \times 3}^{d^{-1}}\right)}_{li}\right)
\label{eq:theta_invariant}
\end{equation}  
where $h_{3 \times 3}^{d^{-1}}$ is the inverse of the $3 \times 3$ sub-matrix corresponding to the yukawa couplings of the light-quarks sector of the down quark mass matrix $M_d$.  In the basis of equation~(\ref{eq:Md_diag_basis}) where this is diagonal, it is easy to check that equation~(\ref{eq:theta_invariant}) reduces to equation~(\ref{eq:theta_higgs_lowM}).  Since equation~(\ref{eq:theta_invariant}) is the trace of a product of matrices all of which transform similarly under unitary transformations in the light $3 \times 3$ sector, it is invariant of the light quark basis.  

The above implies that the contribution to $\bar{\theta}$ due to $\delta M_u$ is  given similarly by~(\ref{eq:theta_invariant}) with $u \leftrightarrow d$ where  $h_{3 \times 3}^{u^{-1}}$ is the inverse of the $3 \times 3$ light quarks yukawa couplings sub-matrix of $M_u$ in equation~(\ref{eq:Mu_Md_diag_basis}).  

An electric dipole moment to the neutron can also be induced due to an external photon line added to the P/CP violating figures~(\ref{oneloopW}) and~(\ref{oneloophiggscharged}) for the first generation up and down quarks. However adding the external photon line does not change the Yukawa matrix structure of these figures, and since both $\bar{\theta}$ and the neutron edm induced directly are one-loop effects, there are no additional bounds we can get in our model by considering these processes.  

We move to the next section where we look at some extensions beyond the minimal model including the possibility of having a singlet so that CP can be broken spontaneously at a high scale.

\section{Spontaneous CP Violation with P even, CP odd singlet.}
\label{sec:SPCP}
Instead of breaking CP softly, we can allow for CP being conserved by the Lagrangian but broken spontaneously. In this case, the direct mass terms $M_i$ are all real and we denote them by $\underline{M}_i$.    We introduce a real CP odd, parity even Higgs singlet $\sigma$ whose non-zero vev will break CP. The singlet introduces additional CP and P conserving Yukawa terms:
\begin{equation}
i \lambda_i \ \sigma \bar{Q}_{iL} {Q'}_{R} \ - \ i \lambda_i \ \sigma \bar{{Q'}}_{L}  Q_{iR} \ + \ h.c
\end{equation}
where $ \lambda_i$ are real. CP is broken when the singlet picks up a parity even real VEV $<\sigma>$. The mass matrix in~(\ref{eq:massmatrix}) remains Hermitian and  has P symmetric CP violating terms with ${M_i} = {\underline{M}_i} + i \lambda_i <\sigma>.$  Note that in this case the Lagrangian conserves both P and CP.  Therefore this is useful in $SO(10)$ motivated left right theories where $C$ can be a symmetry.  

Note also that in this version we can allow soft P violating dimension-2 mass terms without affecting the strong CP solution.  For example $M_R$ and $M_L$ can be different in the terms $M_R^2 {\Delta_R}^\dagger \Delta_R + R \rightarrow L$. But CP would then need to be broken only spontaneously.
\section{Domain Wall Problem} 
When a discrete symmetry like P or CP is broken spontaneously it leads to domains of opposite P or CP  within the universe separated by walls whose energy may dissipate slowly and overclose the universe~\cite{Zeldovich:1974uw}. If the spontaneous breaking occurs higher than the scale of the inflation of the universe (ie $\geq 10^{10} GeV$), then these domain walls are inflated away and pose no problem.  In our case only one of P or CP needs to be spontaneously broken and the other can be broken softly.  For example without a singlet,  P is spontaneously broken while  CP is broken softly and can be at a low scale. While the singlet provides the opportunity to break P softly at a low scale. 

In the first case where there is no singlet introduced, for the Strong CP solution to work it is interesting that CP violation \itshape has \normalfont to be soft and automatically no domain wall problem is associated with it.  On the other hand P violation is spontaneous, and therefore needs Higgs fields whose masses have quadratic divergences from the GUT scale. This then may be the reason why the P violation scale would be higher than the inflation scale.

\section{More Families}
While we have illustrated the case of 3 light and 1 full vector like family in this paper, it is easy to see that the mass matrix with $\ell$ light and $m$ full vector-like families is also Hermitian and solves the strong CP problem.   The mass matrix corresponding to equation~(\ref{eq:massmatrix}) is:

\begin{equation}
M_u =
\left(\begin{array}{cc}      
\underline{H}_u v_L & M \\
M^{\dagger}&  \underline{H}'_u v_L 
\end{array}\right)
\label{eq:genmassmatrix}
\end{equation}
where $\underline{H}_u$ and $\underline{H}'_u$ are $n \times n$ and $m \times m$ real symmetric matrices respectively and $M$ is complex $n \times m$ matrix that provides masses to the vector-like quarks and contains CP violating phases. $\ell = n - m$ are the light chiral quark families and $v_L$ is real. 

On diagonalizing the above we can see that iso-singlet member of each vector-like family has the same tree-level mass as the corresponding iso-doublet member in the absence of $v_L$, just like in the case of one full vector-like family.
 
Moreover if  $\underline{H}'_{u,d}$ are diagonal or block diagonal due to some symmetries on the Yukawa couplings of the primed quarks, then the vector-like masses of each block are separately protected and they could form a chain of vector-like families with independent mass scales. 
 
\section{Concluding Remarks}
By introducing a full vector-like quark family we could separate out the P and CP violating sectors of the left-right model. P violation requires the Higgs fields whose VEVs conserve CP.  On the other hand  CP violation happens softly due to direct dimension-3 quark mass terms that do not require the Higgs fields. It is possible to introduce Higgs fields  for CP violation as well, as we showed with a CP odd, parity even singlet. But then they are done such that their VEVs do not break P.   While manifest left-right symmetry has been \emph{assumed} in many a study, it emerges as a consequence of this solution of the strong CP problem.

$\bar{\theta}$ is generated radiatively at the one-loop level where the P and CP violating sectors interact and depends on the Yukawa couplings.  For the case  $M > v_R$, we find that $\bar{\theta}$ is further suppressed by the factor $(v_R/M)^2$ which can  arbitrarily small for large $M$, irrespective of the strength of Yukawa couplings. On the other hand if $M < v_R$ the smallness of $\bar{\theta}$ is directly related to the smallness of Yukawa couplings involving the fourth family and the observed heirarchy in couplings of the existing 3 families.   

In this paper, particularly for the case $M < v_R$, we have been guided by 't Hooft's sense of naturalness that a coupling or term can be naturally small if setting it to zero increases the symmetry of the Lagrangian. This guarantees that radiative corrections to the term are likewise small since they too disapper in the symmetry limit. Since setting  mass-terms $M_i$ to zero increases the symmetry,  the new vector-like quarks can naturally have masses low enough to be observed at the LHC.  Likewise setting Yukawa couplings of any family to zero increases the chiral symmetry of the Lagrangian and the smallness of Yukawa couplings  is thus natural in the sense of 't Hooft.  This in terms implies that $\bar{\theta}$ which depends on small Yukawa couplings is naturally small.  
   
While we have worked using the minimal left-right model the idea basically is to facilitate P and CP violation in different sectors by using vector-like quarks, and is generalizable to parity-based models beyond the $SU(3)_c \times SU(2)_L \times SU(2)_R$ group. If no Higgs fields are introduced for CP violation, then the automatic symmetry protecting the mass of the vector-like family is not broken by any Higgs VEVs and they can  naturally be at lower energy scales including the TeV scale.   If the full vector-like family is found at the TeV or few times TeV scale, it  can carry the signal of parity-based solution to the strong CP problem by having nearly degenerate masses of its iso-doublet and iso-singlet components. Likewise if it emerges out of other groups beyond $SU(3)_c \times SU(2)_L \times SU(2)_R$ it can also carry those signatures in its composition.  For example if there is grand unification there would also be corresponding vector-like leptons. If these quarks are found at the LHC they could give valuable insight into the physics at higher mass scales as well.
\appendix 
\section{An Example}
\label{sec:appendix}
To see that the CKM phase is generated we will  assume without loss of generality that  {$\underline{H}_u$} and  {$\underline{H}_d$} in equation~(\ref{eq:massmatrix}) are real symmetric and  {$\underline{H}_d \equiv {\underline{H}}^D_d$} is diagonal and given by $ diag \{ \underline{h}_{dd},\underline{h}_{ss}, \underline{h}_{bb},\underline{h}^d_{44} \}$.  
For the purposes of this example we will further assume that only two of the dimension-3 mass terms $M_1$ and $M_4$ are non-zero so that there is only one complex phase.   Let $M_1 = M s e^{i\phi}$ and $M_4 = M c$ with $c^2 + s^2 = 1$.  The $4 \times 4$ unitary transformation $U$ such that $U^{\dagger}_{ij}M_j = \delta_{i4} M$  is given by:

\begin{equation}
U = \left(\begin{array}{ccc}      
  c &  0    &  s e^{i \phi}\\       
 0 & I   & 0 \\
-s e^{-i\phi} & 0   &   c
\end{array}\right)
\label{eq:U_4}
\end{equation}
where $I$ is the $2 \times 2$ identity matrix.  The transformed Yukawa matrices are given by  {$H_{u} = U^\dagger {\underline{H}}_{u} U, H_{d} = U^\dagger {\underline{H}}^D_d  U $}.     We can now write down $M_d$ in the transformed basis:
\begin{equation}
M_d = \left(\begin{array}{ccc|cc}      
h_{dd} v_L &  0  & 0   & h_{d4} v_L &0 \\       
 0 & {h}_{ss} v_L & 0  & 0 &0\\
 0& 0  & {h}_{bb} v_L&   0 &0\\
 \hline
h^\star_{d4} v_L& 0  &  0 &   h_{44} v_L &M\\
 0&0&0&M&h'v_L 
\end{array}\right)
\end{equation}
where $h_{d4} =  s c e^{i \phi} \left( \underline{h}_{dd}- \underline{h}^d_{44}\right)$. The diagonal elements can be expressed  in a similar manner in terms of symmetry basis parameters.  Note that the $3 \times 3$ light quark sector of $M_d$ remains diagonal. We can make $h_{d4}$ real through common $U(1)$ rotations of the $4^{th}$ row and column in up and down sectors of the transformed matrices but we will leave it as is in this example.   

We will now look at the light quark sector in $M_u$ and see that there is a non-trivial CKM phase.  

In the transformed basis, $M_u$ has the general form given by equation~(\ref{eq:Mu_Md_diag_basis}) and for this example, its matrix elements in terms of the symmetry basis elements are:
\begin{equation}
\begin{array}{c}
h_{uc} = c \underline{h}_{uc} - s e^{i \phi} \underline{h}_{4c} \\
h_{ut} = c \underline{h}_{ut} - s e^{i \phi} \underline{h}_{4t}, \ \ \ h_{ct} = \underline{h}_{ct}
\end{array}
\end{equation}

To simplify we further set $\underline{h}_{uc} \approx \underline{h}_{4t} \approx 0$ in the above, so that the only phase in the $3 \times 3$ light quark sector is in the matrix element  $h_{uc}= - s e^{i \phi} \underline{h}_{4c}$.  Now it is easy to see that a non-trivial CKM phase has been generated in the $3 \times 3$ sector. $\underline{h}_{4c}, \underline{h}_{ct}$ and $\underline{h}_{ut}$ are $\sim 10^{-(2 \ to \ 3)}$ or so to give the right CKM angles and $s$ and $c$ can have some intermediate values $\sim 0.5$ or so. 

We can also calculate the strong CP phase generated due to $\delta M_d$ for the case $M < v_R$ using equation~(\ref{eq:theta_higgs_lowM}).  There is only one term contributing in this example and that is:
\begin{equation}
\bar{\theta}  = Im \frac{(h_{uu} + h^u_{44})}{16 \pi^2 h_{dd}}  h_{d4} h_{4u} 
\end{equation}
Writing the off-diagonal Yukawas in terms of symmetry basis parameters we separate out the phase dependence explicitly
\begin{equation}
\bar{\theta}  \sim  \left(\frac{1}{16 \pi^2}\right){(h_{uu} + h^u_{44})  (c^2e^{i \phi} - s^2 e^{- i \phi}) \ sc \ \underline{h}_{u4}}
\label{eq:example_s}
\end{equation}
This implies $\underline{h}_{u4} \leq 10^{-3}$ for $h_{uu} + h^u_{44} \sim 10^{-5}$ so that $\bar{\theta} \leq 10^{-10}.$  If $h^u_{44} \sim 10^{-3}$ then $\underline{h}_{u4} \leq 10^{-5}$. If both $\underline{h}_{u4}, h^u_{44} \sim 10^{-5}$, then $\bar{\theta} \sim 10^{-12}$.

If we turn on $\underline{h}_{uc},  \underline{h}_{4t}$ they will be similarly constrained.  Note also the dependence of~(\ref{eq:example_s}) on $sc$. If instead of the up quark, the top quark coupled to the fourth family with a direct mass term $M_3$, there would be a similar equation with $h_{tt}$ instead of $h_{uu}$, $u \leftrightarrow t$ and $s, c, \phi \rightarrow s_3, c_3, \phi_3$  in~(\ref{eq:example_s}).  Since $h_{tt} \sim 1$ this would now imply that $s_3 c_3 \leq 10^{-3}$ so that $\bar{\theta}$ is within bounds. Thus the mixing $s_3 \sim M_3/M$ would need to be small implying a hierarchy in the direct mass terms.        

In a similar manner we can evaluate the contribution to $\bar{\theta}$ due to $\delta M_u$ using equation~(\ref{eq:theta_invariant}).  This will provide similar bounds on the down-sector Yukawa couplings such as $\underline{h}_{d4}, h^d_{44}$ etc.

Since $M$ could be in the range that the LHC explores,   mass-term ratios could be observable parameters like the CKM matrix elements, and we do not need to worry about there being a hierarchy.  In fact what also seems interesting is that without the hierarchy that already exists, it would be hard to transfer the CKM phase without at the same time inducing large $\bar{\theta}$.  This is because if all Yukawa couplings are $\sim 1$ like the top quark, then all mixings would probably have to be small like $s_3$, and this would then suppress the CKM phase.  On the other hand $M_i/M \leq 1$  along with the  hierarchy that already exists in the light quark sector seems to work easily.

The example shows that it is possible to generate the CKM phase in a consistent way keeping $\bar{\theta} \leq 10^{-10}$.  More possibilities can be found by studying the texture zero aspects of the mass matrices in the symmetry basis.  Flavour symmetries can also be used while keeping in mind that VEVs of any new Higgses should not break the symmetry protecting the vector-like quark masses, so that the model remains testable at the LHC and future colliders.

\begin{acknowledgments}
This work would not be possible without the support and solidarity of my family and friends from Association for India's Development.
\end{acknowledgments}

\bibliography{ravibib}

\end{document}